\documentstyle{aa}
\begin{document}

\thesaurus{11(11.07.1)}

\title{Relativistic corrections to fractal analyses of 
the galaxy distribution}

\author{Marie-No\"elle C\'el\'erier \inst{1}
\and Reuben Thieberger \inst{2}}

\institute{D\'epartement d'Astrophysique Relativiste et de 
Cosmologie, Observatoire de Paris-Meudon, 
5 place Jules Janssen, 92195 Meudon C\'edex, France
\and Physics Department, Ben Gurion University, Beer Sheva, 84105 Israel}

\offprints{M.-N. C\'el\'erier}

\mail{Marie-Noelle.Celerier@obspm.fr
\and thieb@bgumail.bgu.ac.il}

\date{Received 2000 / Accepted}

\maketitle
\markboth{M.N.  C\'el\'erier \&\ R. Thieberger:Relativistic corrections 
to fractal analyses of galaxies}
{M.N.  C\'el\'erier \&\ R. Thieberger:Relativistic corrections 
to fractal analyses of galaxies}

\begin{abstract}
The effect of curvature on the results of fractal analyses of the galaxy 
distribution is investigated. We show that, if the universe satisfies the 
criteria of a wide class of parabolic homogeneous models, the observers 
measuring the fractal index with the integrated conditional density 
procedure may use the Hubble formula, without having to allow for 
curvature, out to distances of 600 Mpc, and possibly far beyond. This 
contradicts a previous claim by Ribeiro (1995) that, in the 
Einstein-de Sitter case, relativistic corrections should be taken into 
account at much smaller scales. We state for the class of cosmological 
models under study, and give grounds for conjecture for others, 
that the averaging procedure has a smoothing effect and that, 
therefore, the redshift-distance relation provides an upper limit to the 
relativistic corrections involved in such analyses. \\

\end{abstract}

\section{Introduction}

Standard cosmology is based on the assumption that the universe is 
spatially homogeneous, at least on scales sufficiently large to justify its 
approximation by a Friedmann-Lema\^itre-Robertson-Walker (FLRW) model. 
The high isotropy measured in the cosmic microwave background radiation 
(CMBR) is usually taken as strong evidence in support of this hypothesis. 
Coupled to a global, respectively local, Copernican principle, this 
observation has led to the formulation of the Ehlers-Geren-Sachs 
(Ehlers, Geren, Sachs, 1968), respectively almost Ehlers-Geren-Sachs (Stoeger, 
Maartens, Ellis, 1995), theorems. Although the assumptions underlying these 
theorems have been discussed by some authors (Clarkson \& Barrett, 1999; 
C\'el\'erier, 2000a), they are generally considered as robust support for 
the Cosmological Principle. \\

The structures seen in galaxy catalogues - groups, clusters and superclusters, 
distributed along voids, filaments and walls - are not viewed as 
contradicting this Principle, as the scales on which the universe is 
assumed to be homogeneous are much larger than those subtended by 
these structures. Furthermore, the most probable existence of dark matter 
implies that the luminous matter seen in the galaxies does not represent 
the entire source of gravitational energy. Thus, the high isotropy of the 
CMBR could, in principle, be reconciled with an inhomogeneous luminous 
matter distribution, provided the missing mass for an homogeneous pattern 
would be provided by dark matter.
This important issue, which has been formulated as the problem of finding 
whether ``light traces mass'' is still the subject of a number of 
interesting works (see e.g., Bahcall et al., 2000), but appears far from 
being settled. However, a majority of cosmologists seems to feel more 
confortable with the idea that the structure distribution should 
present a transition towards homogeneity at a more or less large scale. \\

However, the consensus on a homogeneous feature of structures, even on 
very large scales, has never been complete. The debate extends back to Newton 
(pro-homogeneous) versus Kant and Lambert (pro-hierarchical). The 
pro-hierarchical 
approach was reconsidered in a more modern way by Charlier (1908, 1922),
and later developed by de Vaucouleurs (1953, 1970), who exhibited a  
universal density radius power law. Recent ideas on fractals (Mandelbrot,
1982) led to the pioneering work by Pietronero (1987). Since then, 
improvements in observational techniques have increased the number and 
depth of available three dimensional galaxy catalogues. 
A program of systematic analyses of these catalogues has 
led to the claim that one can identify a fractal  distribution up to 150 
$h^{-1}$ Mpc (Sylos Labini, Montuori, Pietronero, 1998). Although these
results are disputed by many authors (see e.g., Kerscher, 1999), it is now 
believed by many others that one has to increase substentially 
the 5 $h^{-1}$ Mpc lower boundry generally accepted in the early eighties 
for the scale of the homogeneity transition (Davis, Peebles, 1983). The 
ongoing sharp controversy about this issue awaits a new generation 
of galaxy catalogues in order to be settled (Murante et al., 1998; 
Martinez, 1999). \\

One of the assumptions retained by almost every author in the field 
is that Euclidean geometry may be used as a reasonable approximation, up 
to the distances reached by the current galaxy catalogues (Peebles, 1980; 
Sylos Labini et al., 1998). This statement has, however, been disputed by 
Ribeiro (1995), who argued that the usual practice of disregarding 
relativistic effects in the study of galaxy distribution could cease to 
be valid beyond much smaller scales than commonly believed. 
Using the Einstein-de Sitter model as a theoretical framework, he 
claimed that this approximation could be misleading for the scales 
probed by the most recent catalogues. \\

The role of curvature in a universe exhibiting fractal structures has also 
been explored by Humphrey, Matravers and Maartens (1998), using, as a toy 
model, an inhomogeneous isotropic bubble matching a FLRW background. 
Their conclusion was that Einstein's equations and the matching conditions 
imply either a nonlinear Hubble law or a very low large-scale density, 
provided no selection, evolution or cosmological constant effect should 
come into play. \\

Since then, there has been a growth in the literature on this issue. The scale 
beyond which the relativistic corrections become non-negligible for the 
power spectrum analyses of redshift surveys has, for instance, been recently 
explored by de Laix and Starkman (1998) and Mansouri and Rahvar (1999). 
These authors also claim that a light-cone effect is noticeable at lower 
redshifts than usually believed. \\

The purpose of the work we present here is to delve deeper into the tests 
of the sensitivity to relativistic corrections of the methods used in 
calculating the distribution of galaxies. We show that Ribeiro's claim, 
applied to the Einstein-de Sitter case, is ill-founded. Extending our study 
to other homogeneous cosmological models, we show that the Euclidean 
approximation can be considered as reasonable up to scales larger than 
currently reached, for a wide class of standard universes, and that, for 
this peculiar class, the redshift-distance relation applying to a given 
cosmological model provides an upper limit to the relativistic corrections 
involved. \\

In fact, we know that the Friedmann relation between redshift and comoving 
distance has little chance of being valid in our clumpy universe (see e.g., 
Kurki-Suonio \& Liang, 1992; Mustapha et al., 1998; C\'el\'erier, 
2000b). However, in this work, we choose, for simplicity, to study the 
effect of the averaging procedure mainly used for the completion of fractal 
analyses on the relativistic corrections, for peculiar (parabolic) FLRW 
models. We show that, in these models, this procedure has actually a 
smoothing effect and we provide grounds to conjecture that this effect 
could apply to any other kind of (more physical) cosmological models. \\

This article is organised as follows. In Sect.2, we give a short 
description of the statistical methods used to quantify the galaxy 
distribution. Sect.3 is devoted to an analytical study of the 
Einstein-de Sitter case and to a discussion of Ribeiro's results. In 
Sect.4, we describe the virtual catalog method employed to complete 
our numerical simulations, which are presented, with their results, in 
Sect.5. A discussion and the conclusion appear in Sect.6. The analytical 
expressions used for the calculations are given in the Appendix. \\

\section{Quantifying the galaxy distribution}

In this section we give a short description of the methods used 
for quantifying the galaxy distribution. The purpose of this digression 
is to make the arguments concerning our main theme clearer. \\

The standard procedure, pioneered by Peebles (1980), was to follow 
the methods used for gases and liquids. The one-body density, $n$, is 
assumed to be constant. The two-body density, denoted by $n^{(2)}({\bf r}'
,{\bf r}'')$, is given by
\begin{equation}
n^{(2)}({\bf r}' ,{\bf r}'')= n^2 g(\| {\bf r}'' - {\bf r}' \|)
\label{n2}
\end{equation}

Here $g(r)$ is the pair distribution function. The main interest lies in 
the deviation of $n^{(2)}$ from independent particle behaviour and so a two 
point correlation function is defined by $\xi (r) = g(r)-1$. According to 
the results obtained by Peebles,
\begin{equation}
\xi (r) = ({r \over r_0})^{-\gamma},  \qquad r<r_0.
\label{n3}
\end{equation}

The value for $r_0$ was given as 5 $h^{-1}$ Mpc. \\

It was pointed out by Pietronero (1987) that the $r_0$ value is a spurious
result due to the assumption that $n$ is a constant. From 
Eq.(\ref{n2}), we see that the density at ${\bf r}''$ under the condition 
that there is a galaxy at ${\bf r}'$ is given by the conditional 
density $n^{(1)} ({\bf r}'') g(\| {\bf r}'' - {\bf r}' \|)$. If all 
galaxies are equivalent, we can choose ${\bf r}' =0$ and then,
\begin{equation}
n^{(1)} ({\bf r}'') g( {\bf r}'')={n^{(2)}(0,{\bf r}'')\over n^{(1)}(0)}.
\label{n4}
\end{equation}

Assuming isotropy (i.e. all directions to be equivalent), the right 
hand side of Eq.(\ref{n4}) depends solely on $\| {\bf r}'' \|$, and we 
can write:
\begin{equation}
n^{(1)} ({\bf r}) g( {\bf r}) = \Gamma (r), \qquad r=\| {\bf r} \|.
\label{n5}
\end{equation}

Here $\Gamma $ is defined by the right hand side of Eq.(\ref{n4}). As it 
has been shown by Pietronero (1987), $\Gamma (r)$ is the physically 
significant expression, rather than $\xi (r) $. If we can assume 
$n^{(1)}$ = const., then we can write a relation between $\xi (r)$ and 
$\Gamma (r)$:
\begin{equation}
\xi (r) = {\Gamma (r) \over n^{(1)} (0)} - 1.
\label{n6}
\end{equation}

Just as in Eq.(\ref{n3}), we can write for $\Gamma$,
\begin{equation}
\Gamma (r) =const. r^{-\gamma_{\Gamma}}.
\label{n7}
\end{equation}

It was pointed out by Thieberger et al. (1990) that the analysis of data 
does not necessarily give the same value for $\gamma$ and $\gamma_{\Gamma}$. 
Taking a set of points from a known fractal and performing these 
calculations showed that the theoretical value corresponds to that 
obtained from $\gamma_{\Gamma}$. \\

To reduce the problem caused by finite data sets, a method of coarse graining 
has been introduced, with the integrated conditional density 
$\Gamma^*$ (Coleman and Pietronero, 1992):

\begin{equation}
\Gamma^*={1\over V}\int_V \Gamma dV.
\label{gamstarl}
\end{equation}

It produces an artificial smoothing of rapidly varying fluctuations, 
but correctly reproduces global properties. The integral over $\Gamma$
is strongly related to the correlation integral (Grassberger and Procaccia, 
1983),

\begin{equation}
C_2(r)= {1\over {N'}}\Sigma_i \left[
{1\over {N-1}} \Sigma_{j\ne i} \Theta(r - |{\bf X}_i - {\bf
X}_j|)\right].
\label{proc}
\end{equation}

$\Theta$ is the Heaviside function. The inner summation is over the whole 
set of $N-1$ galaxies with coordinates ${\bf X}_j$, $j\ne i$, and the 
outer summation is over a subset of $N^{\prime}$ galaxies, taken as 
centers, with coordinates ${\bf X}_i$. By taking only the inner $N^{\prime}$ 
galaxies as centers we allow for the effect of the finiteness of the sample, 
(see e.g. Sylos Labini et al., 1998).  This procedure is also widely used 
in astronomy (Provenzale, 1991).

\section{Analytical study of the Einstein-de Sitter case}

As Ribeiro (1992, 1995; hereafter refered to as R92 and R95) and other 
authors cited therein quite rightly pointed out, galaxy observations 
are carried out along 
our past light cone. In a homogeneous FLRW cosmological model, the 
proper density is constant on hypersurfaces of constant proper time. 
Therefore, as one looks back to higher and higher redshifted galaxies, 
the crossing of the null geodesics through hypersurfaces of constant proper 
time and density implies changes in the observed density and an apparent 
inhomogeneity can be measured. One can thus wonder whether the 
inhomogeneous pattern identified by the fractal analysis of galaxy 
counts is  a mere artefact of this light cone effect. \\

In R95, the magnitude of this effect in an Einstein-de Sitter 
universe is examined. The conclusion is that ``even accepting 
an error margin of 25\% in the measurements of the global density, a 
redshift equal to 0.1 is approximately the deepest scale where we could 
observe a homogeneous distribution of dust in an Einstein-de Sitter 
model'', unimpaired by the light cone effect. Now, such a redshift 
corresponds to a luminosity distance of about 410 Mpc for a value 
of the Hubble constant $H_0=75$ Km s$^{-1}$ Mpc$^{-1}$ and 310 Mpc 
for $H_0=100$ Km s$^{-1}$ Mpc$^{-1}$. This is actually less than the 
depth reached by the ESO Slice Project galaxy survey (Vettolani et al., 
1997, 1998), which recently fueled the controversy about the transition 
towards homogeneity of the luminous matter distribution (Scaramella et 
al., 1998; Joyce et al., 1999). Other surveys aimed at going beyond 
this limit are currently under way. \\

Ribeiro's argument is based on the analysis of the sensitivity of the 
integrated conditional density $\Gamma^*$, Eq.(\ref{gamstarl}), 
relative to the redshift. We show, in the present section, that the 
expression retained by Ribeiro for $\Gamma^*$ has to be modified, which 
results in $\Gamma^*$ being less sensitive to the redshift value than 
claimed in R95, thus impairing its conclusions. \\

In R92, the author establishes expressions for some observational 
quantities, in an Einstein-de Sitter space-time, as a function of the 
radial comoving coordinate $r$, chosen as a parameter along the null 
geodesics. We list below, and give in R95 units, the quantities relevant 
for our purpose. \\

The cumulative number count $N_c(r)$ is the number of sources which lie at 
radial coordinate distances less than $r$, as seen by the observer at 
$r=0$. In R95, it is written as
\begin{equation}
N_c={2r^3\over {9M_G}},
\label{nc}
\end{equation}
where $M_G$ is the average galactic rest mass. \\

The luminosity distance $d_l$ of a source is the distance 
from which the radiating body, if motionless in an Euclidean space, 
would produce an energy flux equal to the one measured by the observer.
It thus verifies
\begin{equation}
{f=}{L\over {4\pi d_l^2}} , 
\label{dl}
\end{equation}
$L$ being the absolute luminosity, i.e. the luminosity in the rest 
frame of the source, and $f$ the measured bolometric flux, i.e. 
integrated over all frequencies by the observer. Its R95 formulation is
\begin{equation}
d_l=9r\left(2\over {3 H_0}\right)^{4\over 3}\left[\left(18\over H_0\right)
^{1\over 3}-r\right]^{-2}.
\label{dleds}
\end{equation}

And the redshift $z$ runs as
\begin{equation}
1+z=\left(18\over H_0\right)^{2\over 3}\left[\left(18\over H_0\right)
^{1\over 3}-r\right]^{-2}.
\label{red}
\end{equation} 

From the definition of $\Gamma$, given in the previous section, one
obtains, in Euclidean space-time, (see also Sylos Labini et al., 1998):
\begin{equation}
\Gamma(r)={{1\over S}{dN_c\over {dr}}},
\label{gam}
\end{equation}
where $S$ is the area of a spherical shell of radius $r$. \\

Arguing that the relevant distance for an analysis performed in a curved 
space-time is $d_l$, Ribeiro defines $\Gamma (d_l)$ as
\begin{equation}
\Gamma (d_l)={{1\over S}{dN_c\over {d(d_l)}}},
\label{gameds}
\end{equation}
where $S$ is presented as being the area of the observed spherical shell, 
of radius $d_l$, and is thus written
\begin{equation}
S(d_l)\equiv 4\pi d_l^2
\label{sdl}
\end{equation}

It is here worth noting that the luminosity distance is the observable 
quantity relevant for {\it radially} measured distances. It is thus 
correct to use this quantity to evaluate an elementary change in the number 
density $N_c$ as measured by the observer on his light cone. But, when 
looking at a cross sectional area $dS$ perpendicular to the light ray and 
subtending a solid angle $d\Omega$, the observer must consider 
the area distance $d_a$, also called ``observer area distance '' by 
Ellis (1971) and ``corrected luminosity distance'' by Kristian and Sachs 
(1966), defined by
\begin{equation}
dS=d_a^2d\Omega,
\label{ad}
\end{equation}
and related to the luminosity distance of an object measured with the same 
redshift by
\begin{equation}
d_l=d_a(1+z)^2
\label{dlad}
\end{equation}

Therefore, the area of the observed spherical shell, at a given redshift 
$z$, used in Eq.(\ref{gameds}) must be written
\begin{equation}
S={{4\pi d_l^2}\over (1+z)^4}.
\label{sdla}
\end{equation}

From Eqs.(\ref{dleds}) and (\ref{red}), we get an expression for 
$d_l$ as a function of the redshift $z$:
\begin{equation}
d_l(z)={2\over H_0}(1+z-\sqrt{1+z})
\label{delz}
\end{equation}

Using Eqs.(\ref{nc}), (\ref{gameds}), (\ref{sdla}) and (\ref{delz}), 
we then derive $\Gamma(z)$:
\begin{equation}
\Gamma (z)={3H_0^2\over {8\pi M_G}}\left(1+z\over {2{\sqrt{1+z}-1}}\right).
\label{gamzl}
\end{equation}

As can be easily seen, by comparing Eqs.(\ref{sdl}) and (\ref{sdla}), 
this expression differs from the one that would proceed from R95 
by a factor $(1+z)^4$. Therefore, its sensitivity to the redshift value 
is much lower. \\

The volume element $dV$, relevant for calculations on the light cone, is
\begin{equation}
dV=Sd(d_l)=4\pi{d_l^2\over (1+z)^4}d(d_l)
\label{dv}
\end{equation}

Substituting Eq.(\ref{delz}) into Eq.(\ref{dv}) and integrating, we 
obtain
\begin{equation}
V={32 \pi\over H_0^3}{(\sqrt{1+z}-1)^3(2z+6\sqrt{1+z}-1)\over 15(1+z)^{
5\over 2}}
\label{ve}
\end{equation}

Substituting this expression into Eq.(\ref{gamstarl}) and using 
Eq.(\ref{gamzl}), we get
\begin{equation}
\Gamma^*(z)={15H_0^2\over {8\pi M_G}} {1+z\over {(2z+6\sqrt{1+z}-1)}}
\label{gamstarz}
\end{equation}
which can be easily compared to Eqs.(19) and (20) in R95. We thus see that, 
contrary to Ribeiro's claim, $\Gamma^*$ differs from the average number 
density $<n>$ if the correct relativistic quantities are retained. It is 
interesting to note that, as $d_l(z)$, $\Gamma^*$ 
is an increasing function of $z$, not a decreasing one, as is Ribeiro's 
$<n>(z)$ function. \\

Furthermore, if we take temporarily $H_0=75$ Km s$^{-1}$ Mpc$^{-1}$ to 
compare 
our results with those proposed in R95, we see that between $z=0$ and 
$z=1$ the increase of our $\Gamma^*$ is only 5\% (instead of a 87.5\% 
decrease for Ribeiro's).  A significant 
departure of $\Gamma^*$ from its value in an Euclidean universe should 
therefore occur at redshifts higher than claimed in R95. Its ensuing 
conclusions thus do not agree with the analysis, which is confirmed by 
the results of our simulations presented below. \\

It is also interesting to note that the 5\% increase 
in our $\Gamma^*$ at $z=1$ is markedly less than the 17\% increase 
of the $d_l/r_E$ ratio, where $r_E=z/H_0$ (in units $c=1$) is the 
Euclidean distance. In an Einstein-de Sitter model of the 
universe, the redshift-distance relation therefore provides a large 
upper-limit for the departure of $\Gamma^*$ from its Euclidean counterpart. 
The simulations presented below show that this result can be generalized 
to a wide class of standard homogeneous models of the universe. \\

\section{The virtual catalog method}

A real catalog gives, for each observed galaxy, a set of four 
quantities: two angles $\theta$ and $\phi$, the redshift $z$ (or, 
equivalently, the recession velocity $v$) and the 
apparent luminosity $f$ (or, equivalently, the apparent magnitude $m$). 
This is usually obtained by measuring galaxies with $f>f_{lim}$ (or, 
equivalently, $m<m_{lim}$), in a region of the sky delimited by a solid 
angle $\Omega$. \\

In an analysis as described in Sect.2, two of the above observables 
are transformed, for each galaxy, into non-observable quantities, 
the Euclidean distance $r_E$ and the absolute luminosity $L$ (or, 
equivalently, the absolute magnitude $M$), using the following equations:
\begin{equation}
r_E={{cz}\over H_0},
\label{eud}
\end{equation}
\begin{equation}
L = 4 \pi  r_E^2 f,
\label{lum}
\end{equation}
or, equivalently, in Mpc:
\begin{equation}
M=m-5log_{10}r_E-25.
\label{abm}
\end{equation}

Two different kinds of galaxy catalogues are currently available. Pencil 
beams are generated by deep surveys of a region of the sky delimited by a 
very narrow solid angle $\Omega$. Much wider distributed samples are 
generally more limited in depth. A peculiar analysis method applies to 
each of these two types of datasets. \\

Owing to their very small orthogonal extension, pencil beam catalogues 
can only be analysed using the radial count $N_c(<r)$ 
from the observer (Joyce et al., 1999). As we stressed in 
Sect.3, the sensitivity of the cumulative number count $N_c$ 
to relativistic corrections is well represented by the 
redshift-luminosity distance relation. In the 
Einstein-de Sitter model, the departure of this relation from its 
Euclidean counterpart is only 17\% at $z=1$, which is less than the 
25\% error margin in the measurements accepted in R95. The 
Euclidean approximation can thus be considered as valid up to higher 
redshifts than $z\sim 0.3$, currently reached in these surveys. 
In the following, we shall 
not be concerned with the study of relativistc corrections applying 
to this method, but shall focus our attention on the 
more accurate one described below. \\

The second  method, in which one averages over a large number of (inner) 
galaxies, allows us to fit, inside the sample volume, spheres with a 
sufficiently large maximum radius, since the $\Gamma$ statistic can only 
be computed up to scales limited by this radius. To avoid selection effects, 
one generally extracts from the data a set of volume-limited (VL) samples, 
each sample containing every galaxy with an absolute luminosity below $L_{VL}$ 
(or, equivalently, with an absolute magnitude beyond $M_{VL}$) and distance 
below $R_{VL}$, such as
\begin{equation}
L_{VL}=4 \pi  R_{VL}^2 f
\label{lumvl}
\end{equation}
or, equivalently, in Mpc:
\begin{equation}
M_{VL}=m-5log_{10}R_{VL}-25.
\label{mvl}
\end{equation}

The fractal analysis is thus applied to the set of VL samples (Coleman, 
Pietronero, 1992; Sylos Labini et al., 1998). Therefore, relativistic 
corrections must be considered twice. First, for the determination of 
distances, and second, for the construction of the volume-limited 
samples extracted from the catalog. As expected, our numerical simulations 
show that the dominating effect is the distance (or light cone) effect. \\

We here propose to approach the study of these relativistic corrections 
from a different point of view, by considering the light cone problem 
stated as follows. \\

A homogeneous universe, with a random galaxy distribution on each constant 
proper time hypersurface, exhibits, when observed on our past light cone, 
a given pattern of structures (see above Sect.3). If, for a chosen FLRW 
model of the universe, we 
can construct, from a random distribution of points, a virtual catalog as 
it could be actually observed, we can apply, to this virtual catalog, 
the above described classical statistical analysis method, in the Euclidean 
approximation, and consider the fractal dimension obtained. The departure 
of this dimension from its homogeneous value, $D=3$, provides a measure of 
the error made when ignoring curvature effects. \\

In this work, we limit ourselves to the study of spatially 
flat FLRW models. Spatial flatness seems to be a not too restrictive 
prescription, as it is compatible with the most recent results of 
CMBR anisotropy measurements (de Bernardis et al., 2000; Balbi et al., 
2000). \\

The line-element in proper time $t$ and comoving coordinates $r, \theta, 
\phi,$ can thus be written
\begin{equation}
ds^2=c^2dt^2 -a^2(t)[dr^2+r^2(d\theta^2+sin^2\theta d\phi^2)]
\label{lel}
\end{equation}
where a(t) is the scale factor which, in the matter dominated area, evolves 
as $t^{2\over 3}$. \\

On each hypersurface $t=const.$, the galaxy distribution is random 
(homogeneous), with an absolute luminosity distribution given by an 
assumed luminosity function $\phi (L)$. Galaxies evolve on world lines 
$(r=const., \theta=const., \phi=const.)$. Thus, if we retain the 
assumption that, at the distances probed by the current surveys, 
galaxy evolution can be ignored, a given galaxy, with comoving 
coordinates $(r, \theta, \phi)$ and absolute luminosity $L$, will keep the 
same values for these quantities on each $t=const.$ slice. \\

In the geometrical optics approximation, the light received from the 
galaxies travels on null geodesics (see e.g Kristian and Sachs, 1966) 
and thus the objects compiled in the catalogues are all 
located on our past light cone. At each measured value of the redshift $z$ 
corresponds an intersection of this light cone with some $t=const.$ 
hypersurface. \\

To reconstruct the catalog which would be observed, in a given FLRW 
universe, by an earth-grounded observer looking on our past light cone 
from the ``here and now'' point $(t_0, r_0=0)$ to a random distribution 
of galaxies, we adopt the following method. \\

First, on the ``now'' hypersurface $t=t_0$, we generate a random 
distribution of points, 
i.e. galaxies, fixing, for each of them, values for $r<r_{max}$, 
$\theta<\theta_{max}$, $\phi<\phi_{max}$ and $L>L_{min}$. Contrary to 
the producers of real catalogues, we are not 
limited in angular observation, so we can take the angular 
maximum values as large as we wish, up to a complete sky. The 
absolute luminosity distribution is given by the luminosity function 
retained, see Sect.5. This $t=t_0$ hypersurface corresponds to a 
spatial section of a universe which has evolved from different sections 
$t<t_0$ up to $t=t_0$, to reach this random configuration. \\

Then, for each galaxy with a radial comoving coordinate $r$ and absolute 
luminosity $L$, as first generated on the $t=t_0$ hypersurface, we 
calculate the redshift $z$ and apparent luminosity $f$, as they would 
be measured on the observer's past light cone, if the chosen FLRW model was 
to constitute a reliable description of our universe. For this purpose, 
we use the definition of the luminosity distance:
\begin{equation}
f={L\over {4\pi d_l^2}},
\label{lddef}
\end{equation}
where we substitute the expression of $d_l$ as a function of $r$ 
relevant in the given FLRW model. The redshift $z$ also proceeds from its 
expression as a function of $r$ in this model (see the Appendix). \\

We thus obtain the virtual catalog corresponding to a homogeneous galaxy 
distribution in this model, as a set of quadruplets $(f,z,\theta, \phi)$, 
each corresponding to a virtually measured galaxy. \\

Now, we can perform a classical statistical analysis of this virtual 
catalog, as described in Sect.2. We can thus obtain, varying the 
depth of the data sample, different values for the calculated fractal 
dimension $D$. 
The departure of $D$ from its homogeneous value, $D=3$, gives the measure 
of the error made when ignoring curvature effects, at the scale probed by 
the depth of the retained sample, provided the given cosmological model 
can be considered as a good approximation of the observed universe.\\

\section{Numerical simulations}

\subsection{The method of calculation}

We start with random ($x,y,z$). Next, we discard all the points 
for which R is larger than $R_{max} $. Then, we assign to each point a 
``Luminosity'', $L$, obtained from the distribution:
\begin{equation}
P(L)=A*\exp{(-L)} / L.
\label{pl}
\end{equation}

We introduce a cut-off at low luminosities (we chose $L=0.05$) and at high 
luminosities (we chose $L=10.$). From $(L,x,y,z)$ we construct, as 
explained in Sect.4 and in the Appendix, $(f,z,\phi , \theta )$. 
For the Einstein-de Sitter case, we use Eqs.(\ref{ludeds}), (\ref{rededs}) 
and (\ref{lddef}). For the parabolic homogeneous models with a non-zero 
cosmological constant, we use Eqs. (\ref{zedph}), (\ref{dlrph}) and 
(\ref{lddef}). Our choice of the parameters are

$H_0=65$ Km s$^{-1}$ Mpc$^{-1}$

$\Omega_M = 0.3$

$\Omega_{\Lambda } = 0.7.$

Our next stage is to obtain the dimension. We use for that purpose Eq.
(\ref{proc}), which is essentially like using $\Gamma ^*$. As is 
mentioned in Sect.2, we have an inner region at a distance larger 
than some ${\cal D}_{max}$ from the border of the total sample. 
We now calculate distance and luminosity classically, using the Euclidean 
formulae:
\begin{equation}
r_E ={c z \over H_0}, 
\label{rc}
\end{equation}
\begin{equation}
L_E =L {r_E^2 \over {d_l ^2}}.
\label{Lc}
\end{equation}
Our limited sample requires $L_E > L_{min} $. \\

To understand better our results we perform an additional type of 
calculation, considering just two points of the inner region, closest to 
the center and farthest from the center. For each point we can check 
the local density in addition to the dimension in that region.

\subsection{Results}

For each of the three cosmological models studied, we performed 
calculations for six different sets of random numbers and took 
their average. By using these different six catalogues for each 
cosmological model, we could obtain errors on our results. \\

We performed the calculations for different $L_{min}$. For convenience 
we used a parameter $L_M=const*L_{min}$. \\

We then compared our results to those we obtained when ignoring the 
relativistic corrections for the same random numbers. We compared the 
dimensions obtained in both cases using a method developed by Benzi et 
al. (1995). The advantage of this method is that part of systematic errors, 
like border problems, are less pronounced. \\

We checked the case ${\cal D}_{max} = 200$ Mpc. Comparing it to the non-
relativistic case, no significant difference could be detected. \\

Checking the differences for the closest and farthest inner point, we 
obtained that, for the non-relativistic case, 33,162 points were inside the 
sphere, when taking the closest point ($R_{min} = 550$ Mpc). For the 
farthest point ($R_{max} = 1100$ Mpc), we obtained, in the same 
sphere, 33,152 points. This was expected as we chose a homogeneous sample. \\

The results are presented in Table 1. We denote by M1 and M2 the number 
of points in the two spheres of radius ${\cal D}_{max}$, whose centers 
are $R_{min}$ and $R_{max}$ respectively, and by $P_{RET}$, the 
percentage of points retained in each sample, i.e. for three values 
of $L_{min}$. We obtain definitely an influence on the average density, 
depending on the specific model. For the three models, even though this 
value, ($M1/M2$), 
is quite appreciable, still this does not affect the 
dimension within a ${\cal D}_{max} =200 $ Mpc radius. For all the cases, 
we obtain: $D = 3.00 \pm 0.03$ . \\

\begin{table}
\caption{Density results for the three models}
\begin{tabular}{ccccc}
\hline
\ Model$\#$&$P_{RET}$&$M1/M2$&$D(R_{min})$&$D(R_{max})$ \\
\hline
1&100&1.35 $\pm $ 0.02&3.01 $\pm $ 0.04&2.97 $\pm $ 0.03\\ 
2&100&1.12 $\pm $ 0.01&3.00 $\pm $ 0.03&3.00 $\pm $ 0.01\\ 
3&100&1.00 $\pm $ 0.01&3.00 $\pm $ 0.02&3.00 $\pm $ 0.01\\
1&25&1.41 $\pm $ 0.02&2.99 $\pm $ 0.03&2.99 $\pm $ 0.02\\ 
2&25&1.29 $\pm $0.01&2.98 $\pm $ 0.03&3.01 $\pm $ 0.01\\
3&25&1.22 $\pm $0.01&3.00 $\pm $ 0.02&2.97 $\pm $ 0.03\\
1&8&1.46 $\pm $ 0.02&2.99 $\pm $ 0.03&3.01 $\pm $ 0.02\\
2&8&1.42 $\pm $ 0.02&2.97 $\pm $ 0.02&2.99 $\pm $ 0.04\\
3&8&1.40 $\pm $ 0.02&2.99 $\pm $ 0.02&3.01 $\pm $ 0.03\\
\hline

\end{tabular}

-- Model $\# 1$ is the Einstein de Sitter Model.

-- Model $\# 2$ is the Parabolic Homogenous Model with $\Omega_M =0.3$, 
$\Omega_{\Lambda} = 0.7$.

-- Model $\# 3$ is the Parabolic Homogenous Model with $\Omega_M =0.01$, 
$\Omega_{\Lambda} = 0.99$\\

\end{table} 

Our next purpose was to check to what extend the dimension diminishes 
when we go to larger distances. We calculated the case where we consider 
the whole sphere around our galaxy and take for the inner region a midrange 
of the radius so that the centers in that region are farther than 
${\cal D}_{max}$ 
from the borders. The calculation was performed just for one case, the 
Einstein-de Sitter model, and $L_{min}$ was chosen so that 42 $\%$ of the 
galaxies where retained, on average. We averaged over six sets of 
120,000 points, chosen randomly, and over all the abovementioned 
points of the central region (about 12,000), and checked the 
decrease in the dimension when averaging over the ranges 150 -300 Mpc 
versus 450-600 Mpc. We obtain that the dimension slightly decreases by 
$1.6 \pm 0.2 \%$, which is consistent with our other calculations. \\

To compare these weak discrepancies with the light cone effect on the 
luminosity-distance relation, we note that \\
- for the Einstein-de Sitter case, at ${\cal D}_{max} = 200$ Mpc, the 
contribution of the second order term in the power series expansion of 
$d_l$ is a little more than 1\% of the total in Eq.(\ref{dlexeds}), but 
at ${\cal D}_{max} = 600$ Mpc, it is more than 3\%. \\
- for the case $\Omega_M =0.3$ and $\Omega_{\Lambda} = 0.7$, at 
${\cal D}_{max} = 200$ Mpc, this contribution is about 3.2\% of the 
total in Eq.(\ref{dlexpar}). \\
- for $\Omega_M =0.01$ and $\Omega_{\Lambda} = 0.99$, at 
${\cal D}_{max} = 200$ Mpc, it is more than 4\%. \\

We thus see that, except for the Einstein-de Sitter case at 
${\cal D}_{max} = 200$ Mpc, this effect is stronger on $d_l$ than on $D$. 
Furthermore, as, at ${\cal D}_{max} = 200$ Mpc, this effect 
on the fractal dimension $D$ seems to be model independent, the error 
made by ignoring the relativistic corrections in classical analyses must 
be smaller, at this scale, than the statistical noise of the method. 
Therefore, we can safely conclude that, within the studied class of models, 
the redshift-distance relation can actually be used to provide an upper 
limit to the relativistic corrections needed to complete this kind of 
analysis. \\

Furthermore, as Eq.(\ref{dlad}) provides a general relation 
between $d_l$ and the area distance $d_a$, the smoothing $(1+z)^4$ term 
appears in the calculation of $\Gamma^*$ whatever cosmological 
model is retained to represent the observed universe. We are thus inclined 
to suggest that the above property could apply to any model 
and that the averaging procedure has actually some general smoothing 
result.

\section{Discussion and conclusion}

In the present article, we have studied the effect of curvature on the 
results of fractal analyses of galaxy distribution. \\

First, we explored analytically the Einstein-de Sitter model and found 
that, contrary to the claim in R95, the sensitivity of the integrated 
conditional density $\Gamma^*$ to the redshift is far less than the 
sensitivity of the luminosity distance. Therefore, in this model, the 
redshift-distance relation can be used to derive an upper limit to the 
relativistic corrections applying to classical fractal analyses. \\

Then, we enlarged our study to a numerical simulation of a wide class of 
parabolic homogeneous models. To resolve 
every ambiguity which could arise from a mixing of classical and general 
relativistic notions of distances, surfaces, volumes and averagings, we 
calculated the error which could be made by an observer, immersed in a 
FLRW universe, and analysing, with Euclidean statistical tools, an 
intrinsic homogeneous pointlike galaxy distribution, as measured on his 
past light cone. We have shown that, in these models and up to distances 
far beyond those probed by the current 
analyses, the curvature effect is negligible compared to the 
measurement uncertainties. For $H_0=$ 65 km s$^{-1}$ Mpc$^{-1}$, the three 
tested models exhibit, at the scale of 200 Mpc, a discrepancy for 
the numerically calculated fractal dimension of on the order of 1\%, and a 
2\% limit still holds for the Einstein-de Sitter case at scales on the order 
of 450-600 Mpc. \\

In fact, contrary to the author's claim in R95, the procedure averaging 
over many points, far from increasing the curvature effect, has some 
smoothing results, and therefore this effect is weaker on the fractal 
dimension than on 
e.g. luminosity distances. We have actually proved this property for the 
class of universes under study and have given grounds to suppose that 
it can apply to any other cosmological model. \\

Therfore, as regards the current and forthcomimg results which could 
be derived from Euclidean statistical analyses of galaxy data, the 
relativistic corrections could be consistently ignored, up to the 
abovementioned scales, and probably far beyond, in every universe for which 
the dependence of the luminosity distance on the redshift does not too 
widely depart from the Hubble law at the probed scales. \\

We want to mention, for completeness, that the curvature correction is not 
the only bias that can impair the results of such analyses. Evolution 
effects and, mainly, K-corrections are known to generate other kinds of 
errors, but are beyond the scope of the present work. For a discussion of 
the impact of K-corrections on such an issue, we refer the interested 
reader to Scaramella et al. (1998) and Joyce et al. (1999). \\

{\it Aknowledgements.} Part of this work was performed while one of us 
(R.T.) was on a sabbatical leave at the Observatoire de Paris-Meudon. 
He wishes to thank Brandon Carter and Nathalie Deruelle for their 
hospitality. The authors want to thank also the referee for accurate 
remarks which led them to add interesting developments to this work.

\section*{Appendix A: The FLRW models tested}

We give, in this appendix, the expressions for $d_l$ and $z$ retained 
for the conversion of the $(L,r,\theta, \phi)$ data into the 
$(f,z,\theta, \phi)$ quadruplets, in the FLRW models chosen to perform 
the numerical simulations described in Sect.5. We have limited these 
simulations to three examples of the spatially 
flat subclass: the Einstein-de Sitter case, with $\Omega_M=1$ and 
$\Omega_\Lambda=0$, currently the most popular model with 
$\Omega_M=0.3$ and $\Omega_\Lambda=0.7$ and a low matter density case, 
with $\Omega_M=0.01$ and $\Omega_\Lambda=0.99$. As it can be easily seen 
from a comparison of the $d_l$ expansions in powers of $z$ calculated 
below, the low matter density case exhibits the 
larger second order term correction for ``realistic'' flat models with 
$0<\Omega_M \leq 1$. It can thus been considered as a limiting case for the 
study presented here. \\

For the Einstein-de Sitter case, we use Ribeiro's relations in R95, 
corrected for consistency from a dimensional point of view:
\begin{equation}
d_l=r\left(1-{{H_0}\over {2 c}} r\right)^{-2},
\label{ludeds}
\end{equation}
\begin{equation}
1+z=\left(1-{{H_0}\over {2 c}} r\right)^{-2}.
\label{rededs}
\end{equation}
The power series expansion of $d_l$ thus reads
\begin{equation}
d_l={cz\over H_0}+{cz^2\over {4H_0}} +{\cal O}(z^3).
\label{dlexeds}
\end{equation}

\bigskip

For the parabolic homogeneous model with a non-zero cosmological constant, 
the following expressions can be found in text books (see e.g. Carroll, 
Press, Turner, 1992):
\begin{equation}
d_l=(1+z)r,
\label{lumph}
\end{equation}
\begin{eqnarray}
d_l &=& {c(1+z)\over H_0}  \nonumber \\
\int^z_0[(1 &+& z')^2(1+\Omega _M z')
-z'(2+z')\Omega _\Lambda]^{-{1\over 2}}dz' .
\label{dlph}
\end{eqnarray}

It is easy to derive an exact expression for $r$ from 
Eqs.(\ref{lumph}) and (\ref{dlph}):
\begin{equation}
r={c\over H_0}\int^z_0[(1+z')^2(1+\Omega _M z')
-z'(2+z')\Omega _\Lambda]^{-{1\over 2}}dz',
\label{rph}
\end{equation}

For small $z$, we can Taylor expand the right-hand side of Eq.(\ref{rph}) 
and invert to obtain
\begin{equation}
z={2\over {2 \Omega_\Lambda-\Omega_M-2}}\biggl(\sqrt{1+{H_0\over c}
(2 \Omega_\Lambda-\Omega_M-2)r}-1 \biggr).
\label{zedph}
\end{equation}

Substituting Eq.(\ref{zedph}) into Eq.(\ref{lumph}), we get $d_l$ as a 
function of $r$:
\begin{equation}
d_l= r\left[1+{2\over {2 \Omega_\Lambda-\Omega_M-2}}
\biggl(\sqrt{1+{H_0\over c}
(2 \Omega_\Lambda-\Omega_M-2)r}-1 \biggr)\right],
\label{dlrph}
\end{equation}
and can derive the power series expansion of $d_l$:
\begin{equation}
d_l={cz\over H_0}+{c\over {4H_0}}(2-\Omega_M+2\Omega_{\Lambda})z^2
+{\cal O}(z^3).
\label{dlexpar}
\end{equation}

\end{document}